\documentclass[manuscript]{aastex}
\usepackage{graphicx}
\usepackage{color}
\begin{document}

\shorttitle{Multi-wavelength Emission .....}
 \shortauthors{Cheng et al.}

\title{Multi-wavelength Emission from the Fermi Bubble I. Stochastic acceleration from Background Plasma.}
\author{K. S. Cheng$^{1}$, D. O. Chernyshov$^{1,2}$, V. A. Dogiel$^{1,2,3}$,  and C. M.
Ko$^{4}$}
\affil{$^1$Department of Physics,
University of Hong Kong, Pokfulam Road, Hong Kong, China}
\affil{$^2$I.E.Tamm Theoretical Physics Division of P.N.Lebedev
Institute of Physics, Leninskii pr. 53, 119991 Moscow, Russia}
\affil{$^3$Moscow Institute of Physics and Technology (State University), 9, Institutsky lane, Dolgoprudny, 141707, Russia}
\affil{$^4$Institute of Astronomy, Department of Physics and Center for Complex Systems, National Central University,
Jhongli, Taiwan}



\altaffiltext{0}{........}

\begin{abstract}
We analyse processes of electron acceleration in
the Fermi Bubbles in order to define parameters and restrictions
of the models, which are suggested for the origin of these giant
radio and gamma-ray structures. In the case of leptonic origin of
the nonthermal radiation from the Bubbles, these electrons should
be produced somehow in-situ because of relatively short lifetime of
high energy electrons, which lose their energy by synchrotron and
inverse Compton processes. It has been suggested that
electrons in Bubbles may be accelerated by shocks produced by tidal disruption of star accreting onto the central black hole or a process of re-acceleration of electrons ejected by supernova remnants. These processes will be investigated in
subsequent papers. In this paper we focus to study in-situ stochastic
(Fermi) acceleration by a hydromagnetic/supersonic turbulence, in which
electrons can be directly accelerated from the background
plasma. We showed that the acceleration from the
background plasma is able  to explain  the observed
fluxes of radio and gamma-ray emission from the Bubbles but the
range of permitted parameters of the model is strongly restricted.
\end{abstract}


\date{\today}

\maketitle

\section{Introduction}\label{intr}

Recent discovery of a mysterious, diffuse gamma-ray emission
 from the central portion of the Milky Way \citep{dob10,
meng} which are seen as giant features (Fermi Bubbles) elongated
perpendicular to the Galactic plane, was one of the marvelous
discoveries in high energy astrophysics. The gamma-ray spectrum
from the Bubbles is  harder than elsewhere in the Galaxy,
$dN/dE\propto E^{-2}$. This gamma-ray excess in the bubble region
correlates with the earlier discovered   so-called microwave
haze observed  by the WMAP telescope as described by \citet{fink}
and \citet{dob08}, and with the large scale X-ray emission region
first evidenced by analysing the {\it ROSAT} 1.5 keV data, which
clearly showed the characteristic of a bipolar flow
\citep[see][]{snowden}. It was suggested  that  the {\it ROSAT}
structure resulted from a fast wind that drove a shock into the
halo gas with velocity $\sim 10^8$cm/s. This phenomenon requires
an energy release $\sim 10^{55}$erg at the Galactic Center (GC), which should be
periodic in a time scale of $\sim 10^7$yrs \citep{cohen}.

The WMAP haze was uncovered within the latitude range about
30$^{\circ}$ and was roughly bounded by the bipolar structure of
X-rays. The existence of such haze implies a population of
anomalously hard spectrum electrons toward the GC
\citep{dobler12}.

 Recently, the Planck
collaboration \citep{ade12}  detected a residual diffuse emission
in the range above 30 GHz region  surroundings of the GC whose spatial distribution correlated nicely with the Fermi
Bubbles. At Galactic latitudes $|b| < 30^\circ$, the microwave
haze morphology is consistent with that of the Fermi gamma-ray
 Bubbles (FBs). The
   correlation between these two features   implies that the Bubbles are real and
   their multi-wavelength emissions have a common origin. The derived
spectrum is consistent with the power-law  favoring
 synchrotron radiation from  electrons
with a spectrum  $dN/dE \propto E^{-2.1}$. This  implies also
 a new mechanism for cosmic-ray acceleration in the centre of
our Galaxy. Further analysis of radio
emission from the bubble region provided by \citet{carr13} at 2.3
GHz detected two giant, linearly-polarized radio Lobes,
 emanating from the GC. The Lobes  extend ~$60^\circ$, bear a close
correspondence to the Fermi Bubbles. They concluded that the Lobes
are permeated by strong magnetic field strength up to 15 $\mu$G.

It is necessary to mention that the giant structures emanating
from the centre of our Galaxy are not unique. Even more giant
structures are clearly seen in the direction of Cen-A in GHz radio
\citep{junkes93,feain11}, GeV \citep{yang12} and TeV \citep{aha09}
gamma-ray ranges. Recently \citet{stawarz13} found X-ray features
in the lobe of Cen-A which they interpreted as emission of
relativistic electrons in-situ accelerated in the Cen-A lobes up
to energies $\sim 10$ TeV.  Giant X-ray and radio lobes (bubbles)
were found also  in the galaxies  NGC 3801 \citep{croston}, Mrk 6
\citep{mingo11} and Circinus Galaxy \citep{mingo12}.

The origin of the Bubbles is actively discussed in the literature.
These models include some phenomenological assumptions about
processes of energy release and particle production  in the
Bubbles. Thus the assumed energy release in the GC needed for the
bubble formation ranges from $10^{40}$ erg s$^{-1}$  supplied by
star formation regions as assumed by \citet{crock11} to a
hypothetical scenario of  a single accretion with the released
energy about $10^{56}$ erg \citep[see e.g.][]{guo12,Zubo} when a
massive molecular clouds or a star cluster was captured by the
central black hole ten million years ago.

Different mechanisms of gamma-ray production in the Bubbles are
suggested to explain the observed flux from the Bubbles.
 Thus, \cite{crock11}
and \cite{Zubo}  suggested the hadronic origin of gamma-ray
emission from the Bubbles, when gamma-ray photons are produced by
collisions of relativistic protons with that of the background
gas. Alternatively, these gamma-rays can be produced by the
inverse Compton scattering of relativistic electrons on background
photons (leptonic model) and the same electrons generate radio and
microwave emission from the Bubbles via synchrotron  \citep[see
e.g.][]{meng}. There may be several sources (processes) which
generate relativistic electrons in the Bubbles:
\begin{itemize}
\item In-situ stochastic acceleration by MHD-turbulence nearby the Bubble
surface \citep{mertsch}.
\item Acceleration by shocks originated from repeated tidal disruption of stars
captured by the SMBH at the GC \citep{cheng}.
\item Acceleration within jets near the GC about $\sim 10^6$ yr ago, and  subsequent electron transfer
into the bubble by convective flows \citep{guo12,zweibel}.
\end{itemize}

The goal of theoretical models is to explain a number of emission
parameters  which have several remarkable features in the
Bubbles \citep[see][]{meng,dobler12, ade12}:
\begin{enumerate}
\item The structures are symmetrically elongated in the direction
perpendicular to the Galactic Plane.
\item Spectra of radio emission from the Bubbles  are harder than
anywhere in the Galaxy, and the assumed spectrum of electrons is a
power-law, $\propto E^{-2}$.
\item The spatial distribution of emission in the Bubbles shows sharp edges of
the Bubbles.
\item The surface emissivity is almost uniform inside the Bubbles although findings of \cite{hoop}
might indicate that some features of the gamma-ray spectrum at latitudes
$|b|\leq 20^\circ$ could be interpreted as a contribution from the
dark matter annihilation nearby the GC.
\end{enumerate}

Almost all  FB models on the  spectra of nonthermal emission depend on  free parameters  that allows
more or less easily to reproduce the data derived from observations.  We deem that these parameters can be restricted
 quantitatively, if we estimate how many high energy particles  can be generated by this or that mechanism of acceleration.
 The number of high energy particles has not been estimated previously, although this parameter gives a strong restriction on
 acceleration processes as we intend to show from our investigations.

Below we analyse leptonic models of gamma-rays from the FB in which gamma-rays are generated by inverse Compton.
Here we do not suggest a new model of acceleration but instead our goal is to understand whether the existing models are able to provide enough emitting particles.  We analyse this aspect of stochastic acceleration and shock acceleration models in this and subsequent papers.

The stochastic acceleration can be provided by interaction of
charged particles with a hydromagnetic turbulence which is excited
in the halo by jets \citep[see e.g.][]{Zubo} or by a shock
\citep[as assumed by][]{mertsch}. Alternatively this acceleration
is provided by interaction of particles  with a supersonic
turbulence (shocks) which arises from tidal disruptions of stars
captured by the SMBH at the GC as proposed by
\citet{cheng12}. To provide seeds for stochastic acceleration in
the Bubbles, there are no other evident sources of electrons
except those from the background plasma or those injected from the
Galactic plane by sources such as supernova remnants (SNRs),
pulsars, jets, etc., or from $p-p$ collisions in the halo (secondary
electrons). We aim to define  model restrictions  for different
processes of particle acceleration in the Bubbles. In particular
the goal of our analysis is to define whether the processes of
stochastic or mulitple shock acceleration are able to accelerate
electrons up to high energies and at what conditions they provide
relativistic electrons in the Bubble with the required density and
spectrum.

We start   from the case of stochastic acceleration from a background plasma which has it own specificity.
Particularly, in order to estimate the number of accelerated electrons we should estimate a  flux of thermal electrons  running-away  away into the region of acceleration which is generated by  Coulomb collisions of thermal particles \citep[see for details][]{gur60,dog00}. Therefore, we included into the kinetic equation two additional terms describing the Coulomb scattering. The kinetic equations with terms describing particle injection from a thermal plasma have not been investigated in previous models of the FBs.

\section{In-Situ Acceleration from a Background Plasma - General Remarks.}\label{approx}

The kinetic equation for the  distribution function of
electrons, $f(p,t)$, in the case of in-situ acceleration has the
form
\begin{equation}
 {{\partial f}\over{\partial t}}+{1\over p^2}{\partial\over{\partial
 p}}{p^2\left[\left(\frac{dp}{dt}\right)_C f - \left\{D_C(p)+D_F(p)\right\}{{\partial f}\over{\partial
 p}}\right]}+\frac{f}{\tau}=0\,,
 \label{eq_nr}
\end{equation}
where the dimensionless momentum $p$ is in units of $mc$. The distribution function $f$ includes the thermal and
nonthermal components of the particle distribution. Coefficient
$(dp/dt)_C$ describes particle ionization/Coulomb energy losses and
$D_C(p)$ describes  diffusion in the momentum space due to Coulomb
collisions \citep[for equations for these term see][]{ll}. The parameter $\tau$ is the
lifetime of particles in the region of acceleration e.g. due to    escape from there. The stochastic
(Fermi) acceleration is  described as
diffusion in the momentum space with the coefficient $D_F(p)$,
whose value is determined by the frequency of particle collisions
with, e.g. magneto-hydrodynamic fluctuations or shocks.
In the case of scattering by resonant MHD-waves the coefficient has the form \citep[see e.g.][]{ber90}
\begin{equation}
D_F(p)=2p^2\left({v_a\over v}\right)^2\int\limits_0^1 d\mu(1-\mu^2){{
\nu_\mu^+\nu_\mu^-}\over{(\nu_\mu^++\nu_\mu^-)}}
\end{equation}
where
\begin{equation}
\nu^\pm_\mu
\simeq 2\pi^2|\omega_H|{{k_{res}W^\pm (k_{res})}\over H^2}
\label{nu}
\end{equation}
$H$ is a strength of large scale magnetic field, and $W^\pm$ is the power spectrum of MHD-waves propagating
along a magnetic field line in the both directions. Here $\mu$ is the cosine of particle pitch-angle and
\begin{equation}
k_{res}=\left|{{eH}\over{pc^2m\mu}}\right|,~~~~~\omega_H=\frac{eH}{m_ec}
\end{equation}
where $p$ is in $mc$ units.

In the case of stochastic acceleration of electrons by a supersonic turbulence the coefficient of momentum diffusion
 is \citep[see][]{byk93}
\begin{equation}
D_F(p)\sim\frac{u^2}{cl_{sh}}p^2
\end{equation}
where $u$ is the shock velocity, and $l_{sh}$ is the average
separation between two shocks. The acceleration by a supersonic turbulence is possible if the mean path
length of electrons determined by energy losses and spatial diffusion in the intershock medium is larger than the separation between shocks.

The acceleration is effective when the rate of acceleration
exceeds that of losses, i.e. for $p>p_{inj}$
\begin{equation}
 p_{inj}\sim D_F(p)/(dp/dt)_C\,.
 \label{pinj}
 \end{equation}

A naive assumption could be that in the range $p<p_{inj}$ the
spectrum is Maxwellian, and for $p>p_{inj}$ the spectrum is
non-thermal (power-law). However, calculations of the non-thermal
component is non-trivial. As \citet{gur60}  showed, the
acceleration  distorted the equilibrium Maxwellian spectrum of
background particles because of the flux of particles running-away
into the acceleration region. Even in the case when only a small
part of thermal particles are accelerated and the coefficient of
the kinetic equation ($(dp/dt)_C$ and $D_C$) are determined by the
Maxwellian part of the spectrum, that makes Eq. (\ref{eq_nr})
linear, a very broad transfer region between the thermal and
non-thermal parts of the spectrum is generated by the
acceleration.  The calculation showed that the number of
accelerated particles was larger than it followed from trivial
estimates. This linear approximation was used by
\citet{dog00}and \citet{dog07} who interpreted  nonthermal X-ray emission
of galaxy clusters by in-situ accelerated electrons.

However, approximation of \citet{gur60} does not take into account
a backward reaction of accelerated particles on the thermal
component. This effect can be analysed if the coefficients of  Eq.
(\ref{eq_nr}), i.e. $(dp/dt)_C$ and $D_C$, are calculated for the
total distribution function which includes both thermal and
non-thermal components \citep[see for details][]{ll}. Analysis of
the nonlinear version of Eq. (\ref{eq_nr}) was provided by
\citet{wolfe06} and \citet{east08} who showed that  the energy
supplied by sources of stochastic acceleration was quickly
absorbed by the thermal plasma because of the ionization/Coulomb
energy losses of accelerated particles. As a result this
acceleration is accompanied mainly by plasma overheating while a
tail of nonthermal particles is not formed, i.e. the effect of
stochastic acceleration is negligible.

This conclusion was later revised by \citet{chern12} who derived
from analytical and numerical calculations that the efficiency of
stochastic acceleration depended strongly on parameters of
acceleration. For some conditions the conclusion of \citet{wolfe06}
and \citet{east08} holds, i.e., plasma overheating does occur. However, there
are conditions under which the acceleration forms a prominent non-thermal tail while the
background plasma is not overheated.

 \citet{wolfe06} and \citet{chern12}  presented the coefficient $D_F(p)$  in an arbitrary form
as
\begin{equation}
D_F(p) = \alpha p^\varsigma \theta(p-p_0) \mbox{,} \label{pcut}
\end{equation}
where $\alpha$, $\varsigma$ and $p_0$ are arbitrary parameters,
i.e. the acceleration is effective in the momentum range $p>p_0$.
We notice, however, that there are physical reasons for a cut-off at $p_0$, e.g. it may occur in the
MHD-spectrum of turbulence due to processes of wave absorption by accelerated cosmic rays (CRs).
In Appendix \ref{appen} we presented a qualitative estimations for $p_0$. We showed there that for
an appropriate combination of parameters the value of $p_0$ is about 0.2. However, this analysis is
given as an illustration of potential possibility for a cut-off in a low momentum range and cannot
be considered for  comparison with the $p_0$ values derived below in Section \ref{estim} from numerical simulations.

Here we outline the
analysis of in-situ particle acceleration by stochastic acceleration in the Fermi Bubbles. For
details we refer the reader to \citet{chern12}.

The total power supplied by external sources  to electrons
is determined from
Eq. (\ref{eq_nr}) by the following integral
\begin{equation}
{\dot{W}}=-V_{FB}\int\limits_{p_0}^{p_{max}}
\mathcal{E}\frac{\partial}{\partial p}\left[p^2D_F\frac{\partial
f}{\partial p}\right]dp, \label{power}
\end{equation}
where $\mathcal{E}$ is the particle kinetic energy, and
$V_{FB} \sim 10^{67}$cm$^3$ is the volume of the Bubble.

For simplicity we present equations from \citet{chern12} for large enough $p_0$. The
 flux of particles, $S$, running-away to the acceleration region can be presented as
\begin{equation}
S=\alpha(\varsigma+1)p_0^{\varsigma+1}\sqrt{\frac{2}{\pi}}\frac{N}{T}\exp\left(-\frac{\mathcal{E}_0}{T}\right)
\left[1+\frac{\alpha(\varsigma+1)p_0^{\varsigma+1}}{A(p_0^2+1)}\right],
\label{flux}
\end{equation}
where $N$ and $T$ are the  density and the temperature of background
plasma,
\begin{equation}
\mathcal{E}_0 = \sqrt{p_0^2+1}-1\,,
\end{equation}
and
\begin{equation}
A=4\pi r_e^2cN\ln\Lambda.
\end{equation}
Here $r_e$ is the electron radius and $\ln\Lambda$ is the Coulomb logarithm.
The spectrum of nonthermal particles can be presented as
\begin{equation}
f(p) = \bar{f} (\bar{p}/p)^{\varsigma+1},
\label{sp_appr}
\end{equation}
where  $\bar{f}$ is determined from the conditions by fitting
between the thermal and non-thermal components of the spectrum, and $\bar p$ is the momentum value at this boundary.

In the quasi-stationary approximation when variations of $f$ is
quite small the non-thermal component can be presented as
\begin{equation}\label{fII_pwlaw}
f(p) =  -
\frac{S }{\alpha(\varsigma+1)}p^{-\varsigma-1}~~~\mbox{for}~~~p>p_0\,.
\label{spectr}
\end{equation}

The temperature variations due to the run-away flux $S$ and the
Coulomb losses of non-thermal particles can be presented in the simplest case as \citep[for more accurate estimate see][]{chern12}
\begin{equation}
\frac{dT}{dt}=\frac{2S}{3N}\left[\frac{AQ(p_0,\varsigma)}
{\alpha(\varsigma+1)}-\mathcal{E}_0\right] \label{temperature}
\end{equation}
where
\begin{equation}
Q(p_0,\varsigma)=\int\limits_{p_0}^\infty x^{-\varsigma}\sqrt{x^2+1}dx.
\end{equation}

\section{Parameters of the  Model of In-situ Stochastic Acceleration in the
Fermi Bubbles\label{estim}}

Parameters of plasma in the FB are not well-known. Below for
calculations we accept them as they presented in \citet{meng},
namely: the  density $N=10^{-2}$ cm$^{-3}$ and the temperature
$T=2$ keV. Estimations of the magnetic field strength in the FB
ranges from several $\mu$G up to 15 $\mu$G
\citep[see][]{strong10,jones12, carr13}. The goal of these
calculations  is to reproduce the following characteristics of
nonthermal emission from the FB:

\begin{enumerate}
\item The observed spectrum of  gamma-rays has a cut-off at the energy about 100 GeV \citep{meng} which corresponds to the maximum of electron energy about 0.3 TeV \citep[see e.g.][]{cheng};
\item The total
gamma-ray flux  at energies $E>1$ GeV is $F_\gamma\simeq 4 \times
10^{37}$ erg s$^{-1}$, and the spectrum can be approximated by
$E_\gamma^{-2}$ in the range $1-100$ GeV \citep{meng}. This
condition restricts the number of accelerated electrons;
\item  The radio flux from the bubble in the frequency range 20-60
GHz is $(1-5)\times 10^{36}$ erg s$^{-1}$ and the spectral index
of radioemission is about -0.51 \citep[see][]{fink,jones12,ade12};
\item The power of potential sources of energy release in the GC cannot exceed the
value about $10^{40}$ for star formation regions \citep[][]{crock11} and $10^{41}$ erg s$^{-1}$ for tidal processes there   \citep[][]{cheng};
\item Mechanism  of particle acceleration  should
effectively generate nonthermal particles and not to overheat the
plasma \citep{chern12}.
\end{enumerate}

From these conditions  a necessary set  of acceleration parameters,
$\alpha$, $\varsigma$, $\tau$ and $p_0$ can be estimated using numerical simulations.

Assuming that the gamma-ray  emission is produced by the
accelerated relativistic electrons via inverse Compton scattering
we calculated the  intensity of gamma-ray emission along the line of sight ${\bf l}$ from the integral
\begin{equation}
I_\gamma(t,E_\gamma, {\bf l})=\frac{1}{4\pi}\int\limits_{\bf l}d {\bf l}
\int\limits_\epsilon
n(\epsilon,{\bf r})d\epsilon\int\limits_{p}p^2f({\bf r},p,t)\left(\frac{d^2\sigma}{d\epsilon~dp}\right)_{KN}dp\,.
\label{gamma}
\end{equation}
Here $n(\epsilon,r)$ is the spatial distribution of background
photons with the energy $\epsilon$ which was  taken from
\citet{acker},
$\left({d^2\sigma}/d\epsilon~dp\right)_{KN}$ is the Klein-Nishina
cross-section taken from \citet{blum}.

The cut-off in the electron spectrum can be derived from the balance between the acceleration and the energy losses
\begin{equation}
p_{c} =
\left(\frac{\alpha(\varsigma+1)}{\beta}\right)^\frac{1}{3-\varsigma}\,,
\label{cut}
\end{equation}
 if
$\varsigma<3$. Here the synchrotron and inverse Compton losses are
presented as $dp/dt = \beta p^2$. For estimates we took  $H =
5~\mu$G, density of optical photons $w_{op} =$ 1.6 eV/cm$^3$  and
density of IR photons $w_{IR} = $0.33 eV/cm$^3$
\citep[see][]{acker, carr13} that gives $\beta = 1.5\times
10^{-19}$ s$^{-1}$.

 As follows from  calculations of
\citet{cheng}  the electrons should be accelerated in the FB up
to $E_{max} \simeq 0.3$ TeV (condition 1). Then from Eq. (\ref{cut}) we can derive  a function $\alpha(\varsigma)$.

 In the case of stochastic acceleration the coefficients of momentum, $D_F(p)$, and spatial, $K(p)$,
 diffusion are proportional to each other  \citep[see e.g.][]{ber90}
\begin{equation}
K(p)D_F(p) \approx \frac{p^2v^2}{6}
\end{equation}
where $v$ is the characteristic velocity of turbulence. Then the escape time, $\tau$  in Eq. (\ref{eq_nr}) is
\begin{equation}
\tau \approx \frac{L^2}{4K} \approx \frac{3L^2D_F}{2p^2v} = \frac{3\alpha L^2}{2v^2}p^{\varsigma - 2}\,,
\label{tau_pdep}
\end{equation}
where $L<3$ kpc is the size of acceleration region.

The effect of escape is the  steepening of the spectra of
acceleration particles in comparison with the approximation
(\ref{sp_appr}). Then for the known function $\alpha(\varsigma)$ we can derived from Eq. (\ref{eq_nr}) the escape time $\tau(\alpha)$, and thus the spectrum of electrons that generates  the radio flux from the FB as: $I_r\propto \nu^{-0.5}$ (condition 3).

\begin{figure}[h]
\begin{center}
\includegraphics[width=0.5\textwidth]{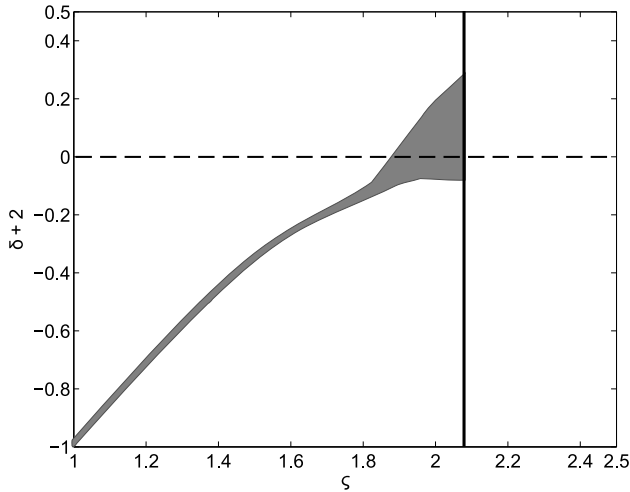}
\end{center}
\caption{ The function $\delta(\varsigma)$ derived from Conditions 2 and 3.} \label{azeta}
\end{figure}

From
Eq. (\ref{gamma}) we can find numerically the value of $\varsigma$ at which the
spectrum of gamma-rays is power-law ($I_\gamma\propto E_\gamma^{\delta}$) with the spectral index of gamma-rays $\delta\simeq -2$ (condition 2). Variations of
$\delta(\varsigma)$ calculated numerically are shown in Fig.
\ref{azeta}.
As one can see from this figure the required value of $\delta$ is
obtained if $\varsigma\simeq 2$. For  other values of $\varsigma$
the solution of (\ref{eq_nr}) does not reproduce the observed
gamma-ray spectrum from the FB. At that, the necessary value of $\tau$ is about
$1.1\times 10^{13}$ s and $\alpha \simeq 1.6\times 10^{-14}$ s$^{-1}$.

The next step of our calculations is to define whether the
acceleration with the derived parameters $\alpha$, $\varsigma$  and $\tau$ can
provide the necessary number of relativistic electrons to
reproduce the observed intensity of the radio and gamma-ray
emission from the FB. As one can see from Eqs. (\ref{flux}) and
(\ref{spectr}) the number of accelerated electrons  depends on the
cut-off momentum  $p_0$, the larger $p_0$, the smaller number of
accelerated particles. However, for the value of $\alpha$ fixed from the cut-off position  in the observed FB gamma-ray flux the maximum value of $\varsigma$ is determined by the density of electrons needed for the observed gamma-ray flux from the FBs and in this respect is independent of other parameters of the model. Just because of this effect the parameter
$\varsigma$ cannot be larger than 2.1 as shown in Fig. \ref{azeta}
by the vertical line. We notice also that although $\varsigma$ is a function of $\alpha$, its estimates from the electron spectrum (see Eq. (\ref{spectr})) or from the cut-off position (see Eq. (\ref{cut})) depends logarithmicaly on $\alpha$, and, roughly, this dependence is neglected in calculations presented in Fig. \ref{azeta}.

\begin{figure}[h]
\begin{center}
\includegraphics[width=1.0\textwidth]{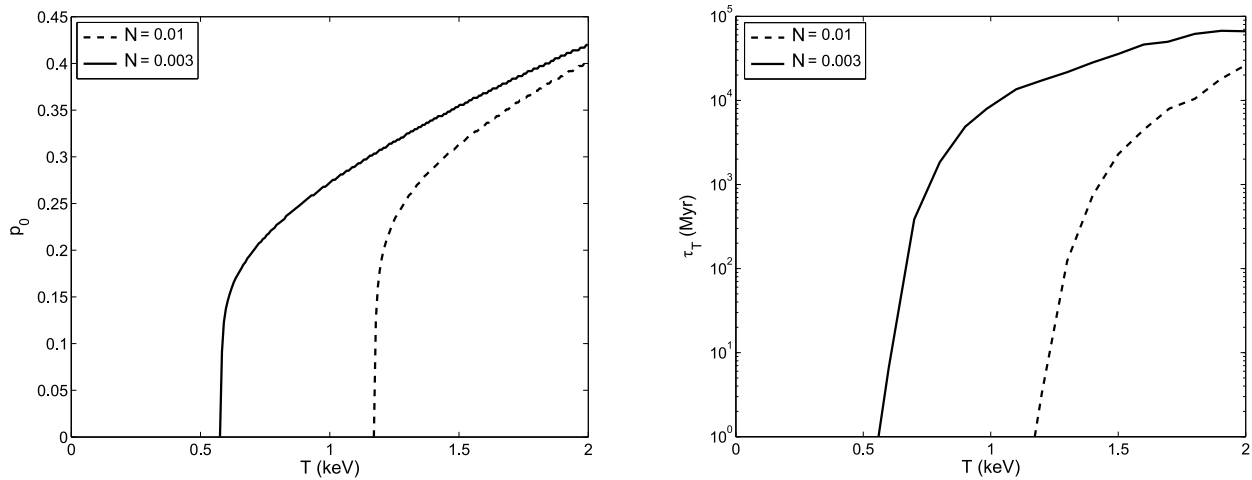}
\end{center}
\caption{a) The function $p_0(T)$ derived from the condition that the number of accelerated electrons is enough to reproduce the observed the nonthermal emission from the FB (left panel); b) The time of plasma heating $\tau_T(T)$ derived from Eq. (\ref{tauT}) (right panel).} \label{p0_and_tau}
\end{figure}

In Fig. \ref{p0_and_tau} (left panel) we showed the  cut-off momentum  $p_0(T)$ at which the number of accelerated electrons is high enough for the observed intensity of gamma-ray emission from the FB (condition 2). The numerical calculations were performed for the two values of the plasma density in the FB: the plasma densities  $N = 10^{-2}$ cm$^{-3}$ (dashed line)
and $N = 3\times 10^{-3}$ cm$^{-3}$ (solid line). One can see that the acceleration can provide enough accelerated electrons if the temperature of background plasma is higher than $\sim 1$ keV.

On the other hand, as it was shown by \citet{chern12} the value of
$p_0$ should not be too small, otherwise the plasma is overheated
by the acceleration particles (condition 5). This condition can be
presented as an inequality
\begin{equation}
\tau_{acc}<\tau_T
\label{ov_T}
\end{equation}
 where the acceleration time $\tau_{acc}\sim 1/\alpha$, and the characteristic time of temperature variation $\tau_T$ is
\begin{equation}
\tau_{T} = \frac{T}{dT/dt}\,,
\label{tauT}
\end{equation}
where temperature variations due to heating by the accelerated electrons are described by Eq. (\ref{temperature}).

From the derived dependence $p_0(T)$ we calculated from Eq. (\ref{tauT}) the time of plasma heating $\tau_T$. The results are shown in Fig. \ref{p0_and_tau} (right panel). The derived acceleration time is about $\tau_{acc}\simeq 2$ Myr. Then the acceleration is possible if $\tau_{acc}<\tau_T$. As it is clear from the figure this condition is realized for  temperatures higher than $\sim 1$ keV.

\begin{figure}[h]
\begin{center}
\includegraphics[width=0.6\textwidth]{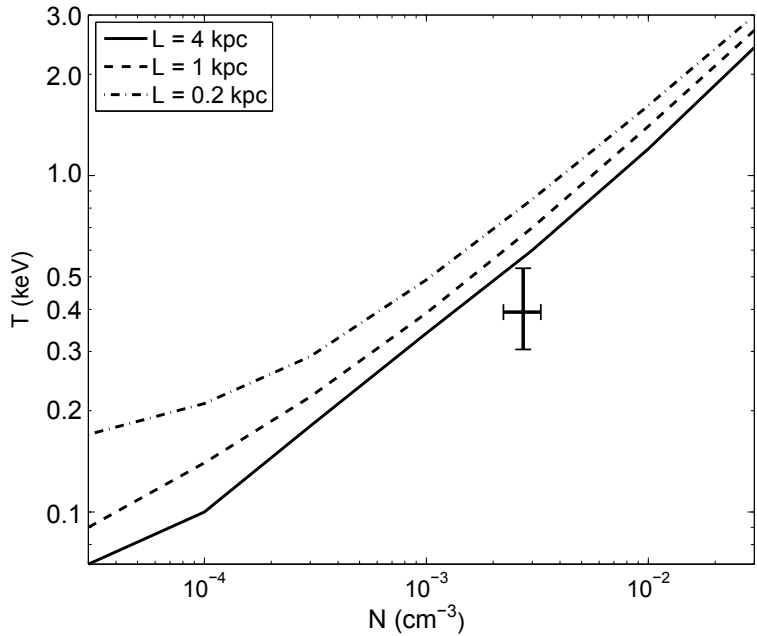}
\end{center}
\caption{The minimum plasma temperature $T$ required for acceleration for a given density of background plasma $N$. Cross marks the observations of the Fermi bubbles by Suzaku \citep{suz_bubble}.} \label{Tmin}
\end{figure}

These conclusions about the plasma temperature derived from the conditions of plasma overheating and of shortage of
accelerated electrons are illustrated  in Fig. \ref{Tmin} where the plasma temperature, $T$, required for  acceleration,  is shown as a function of the plasma density, $N$.  The functions $T(N)$ was derived for different thicknesses of the acceleration region $L$. The values along the curves satisfy the data $N = 10^{-2}$ cm$^{-3}$, T = 2 keV obtainded by \citet{meng}. We also placed in the figure the data obtained by Suzaku \citep[see][]{suz_bubble} for the FB region. It seems to us that there is no serious discrepancy between results of numerical simulations and the observational data. The ROSAT data \citep[see][]{snowden} for the FB region does not differ significantly from that of  Suzaku.

\begin{figure}[h]
\begin{center}
\includegraphics[width=0.6\textwidth]{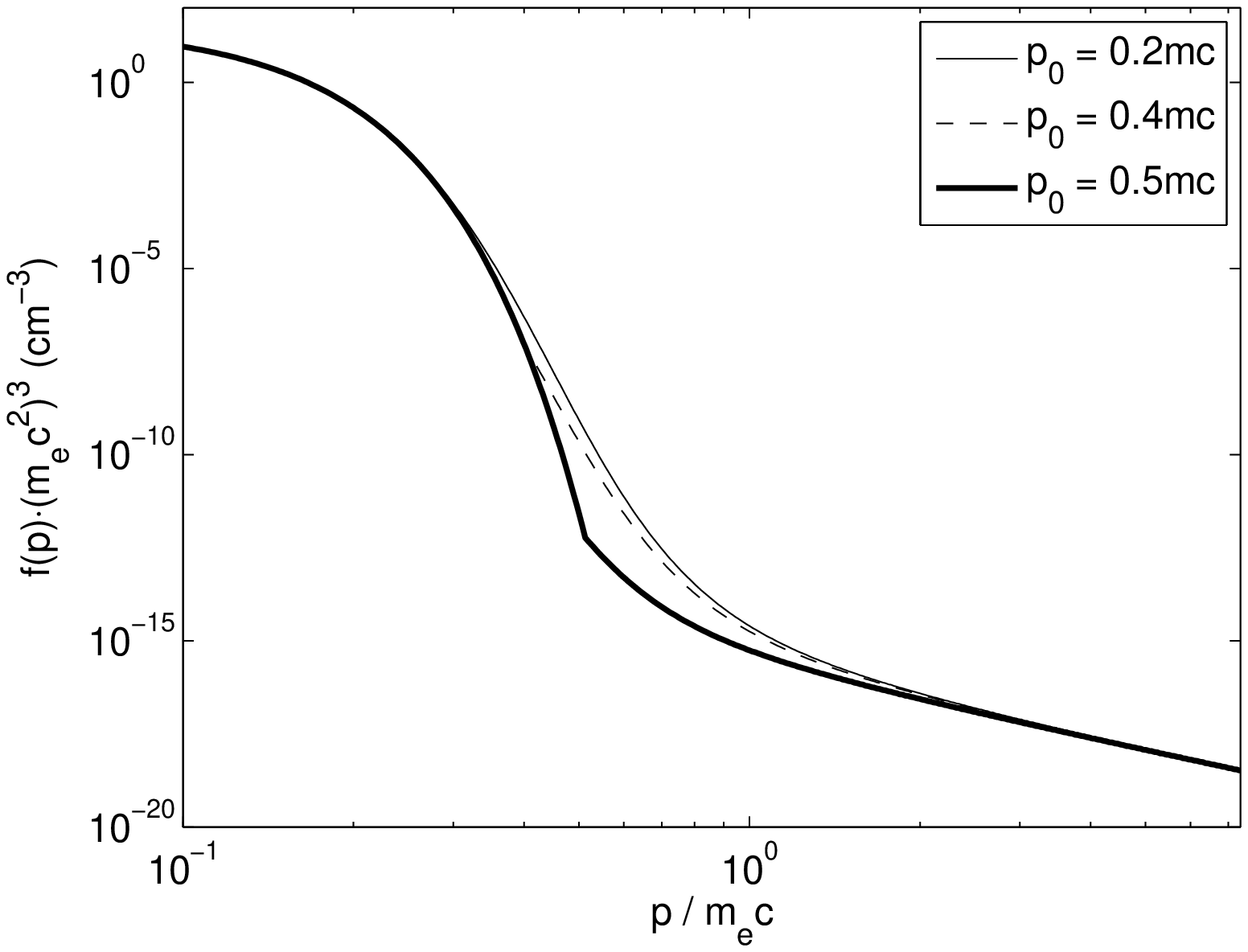}
\end{center}
\caption{The  spectrum of electrons in the Bubbles for various values of $p_0$.}
\label{electr}
\end{figure}

 These two effects of plasma overheating and of shortage of
accelerated electrons are illustrated in  Fig. \ref{electr} where
we showed spectra of accelerated electrons for different $p_0$.

From this figure one can see the reason for restricted values
of $p_0$. For large values of $p_0$ the spectrum of electrons is
below the thick solid line which showed the intensity of electrons
needed for the observed gamma-ray flux. Just for this reason we
have a restriction that $\varsigma\la 2.1$ shown by the vertical
solid line in Fig. \ref{azeta}. For $\varsigma > 2$ the density of accelerated
electrons is  smaller than need for the FB gamma-ray flux. On the
other hand, the stochastic acceleration forms an excess of suprathermal particles nearby the Maxwellian distribution.
For  $p_0<0.2$ this excess is so high  that the electrons from
this excess region heat effectively the plasma. Thus, the thermal pool absorbs the energy supplied by sources  that prevents from
effective acceleration.

\section{Numerical Calculations of Gamma-ray and Radio Emission from the FB}\label{numer}
Here we present results of direct numerical calculations  of
the electron spectrum, gamma-ray and radio emission when the distribution function $f$
is calculated numerically   from Eq.
(\ref{eq_nr}) for the derived parameters of acceleration. Then the FB gamma-ray spectrum is calculated
from Eq. (\ref{gamma}). The expected
radio spectrum at the frequency $\nu$ in the direction ${\bf l}$
is calculated from the following equation \citep[see for details of the
equation][]{syr59,ginz65}
\begin{equation}
I_r(t,\nu, {\bf l})=\frac{1}{4\pi}\int\limits_{\bf l}d {\bf l}
\int\limits_{E}p(\nu,E)F({\bf r},E,t)dE\,.
\label{rad}
\end{equation}
where $E=pc$ for relativistic electrons, $F(E)=p^2f(p)(dp/dE)$, and the function $p(\nu,E)$ is
\begin{equation}
p(\nu,E)=\sqrt{3}\frac{3e^3H_\perp}{mc^2}\frac{\nu}{\nu_c}\int\limits_{\nu/\nu_c}^\infty K_{5/3}(x)dx
\end{equation}
Here $H_\perp$ is the average component of magnetic field perpendicular to  ${\bf l}$ and
\begin{equation}
\nu_c=\frac{3eH_\perp}{4\pi mc}\left(\frac{E}{mc^2}\right)^2
\end{equation}
\begin{figure}[h]
\centering
\includegraphics[width=1.0\textwidth]{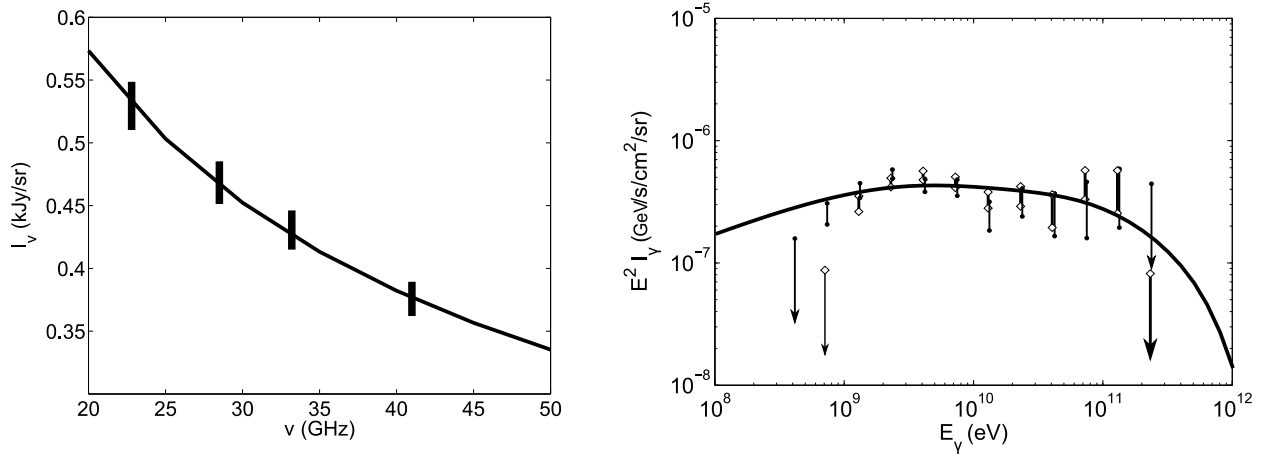}
\caption{Spectrum of radio (left panel) and gamma-ray (right
panel) emission from the FB. The datapoints  were taken from
\citet{meng} and \citet{ade12}. } \label{sp_gamma}
\end{figure}
 For the derived values of $\varsigma$, $\alpha$ and $\tau$ (see previous section) we calculated
numerically the  gamma-ray and radio intensity which are shown in  Fig. \ref{sp_gamma}. At that the needed value of $p_0$ is:
$p_0= 0.34$. 
The results of
calculations coincide nicely with the data.
\begin{figure}[h]
\centering
\includegraphics[width=0.6\textwidth]{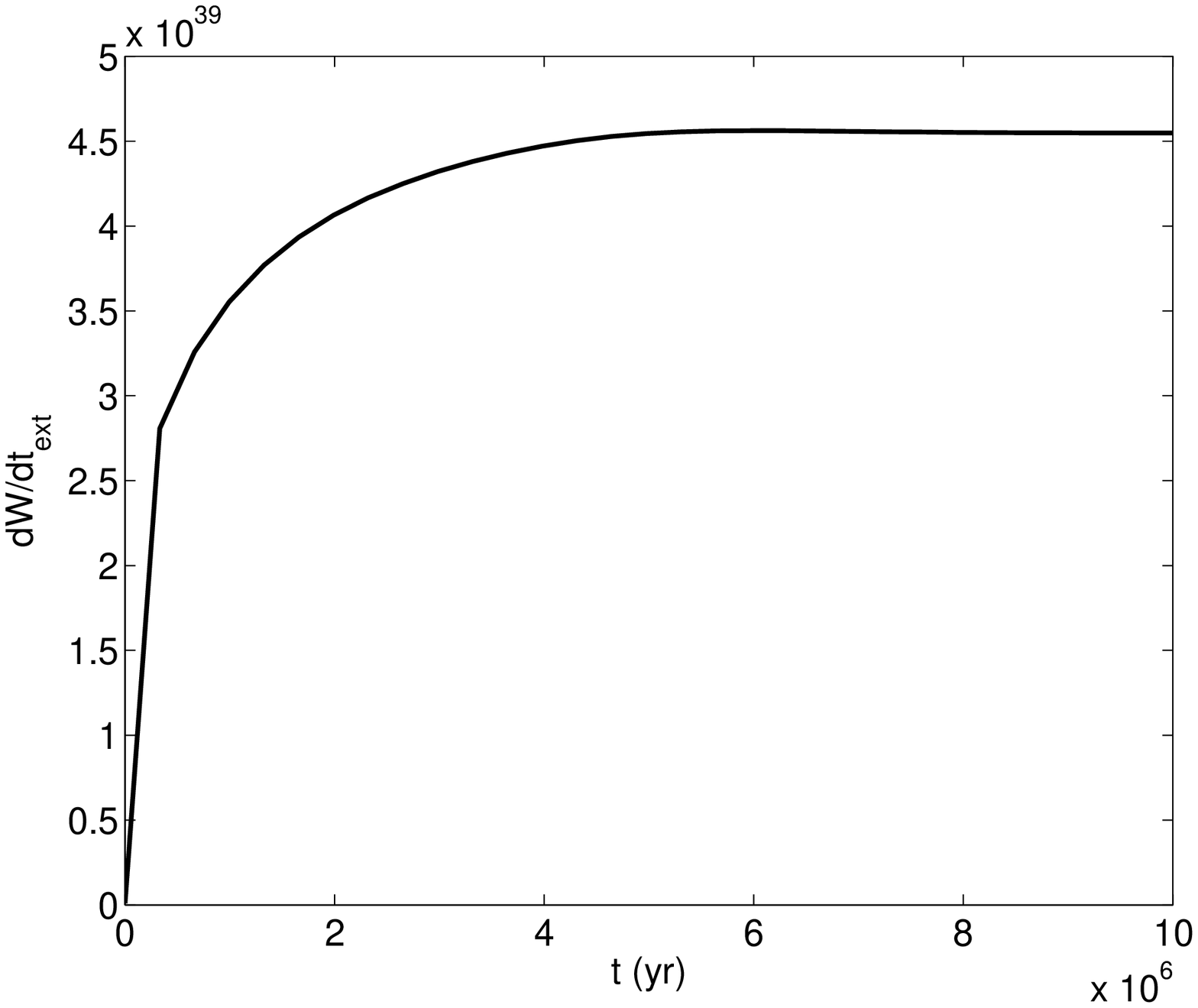}
\caption{Evolution of the power supplied to the system with time.
} \label{fig:wdot}
\end{figure}
Numerical calculations  of the power of external sources as
described by Eq. (\ref{power}) are shown in Fig.\ref{fig:wdot}. It
is accepted here that this sources of acceleration are switched on
at the time $t=0$. As one can see from  the figure the power
reaches its stationary state at $\dot{W}\simeq 4.5\times 10^{39}$
erg s$^{-1}$ for the time $t=4\times 10^6$ yr.

 This value is lower than $10^{40}$ erg s$^{-1}$ as estimated by \citet{crock11} for the energy  release provided by the star formation regions in the GC and also is below $10^{41}$ erg s$^{-1}$  as estimated by \citep{cheng} for star
accretion processes onto the central black hole (condition 4). Thus, we conclude
that
 the measured flux of radio and gamma-rays,  the estimated power of sources and the upper limit of
energy release in the GC are in good agreement with each other in
the model.

\section{Conclusion}\label{concl}
In order to provide high energy electrons responsible for the
electromagnetic radiation (gamma-ray and radio) from the Fermi
Bubbles, we investigated the case of stochastic in-situ acceleration of
electrons from the halo background plasma. The stochastic
acceleration in the FB can be either due to charged particle
interaction with resonant MHD-waves or with a supersonic
turbulence in the FB as it was assumed by  \citet{cheng12}. We
obtained the following conclusions:
\begin{itemize}
\item Two essential assumptions are used in the model: a) the FB gamma-ray emission is  produced by the
inverse Compton scattering of relativistic electrons on the
background Galactic (IR and optical) and  relic photons , and b)
these electrons are accelerated by stochastic (Fermi) acceleration
from the background Galactic plasma whose density and temperature
in the FB are accepted as $N=10^{-2}$ cm$^{-3}$ and $T=2$ keV. The process of stochastic acceleration in the FB can be either due to
 particle acceleration with a supersonic turbulence as assumed by \citet{cheng12} or
 by interactions with resonant-MHD waves \citep[see][]{ber90}. The
goal is to define model parameters at which the gamma-ray and
radio emission from the FB can be provided by this acceleration
mechanism.
\item As it is well-known, the process of Fermi acceleration generates very flat (hard)
spectra of particles which are  harder than needed for the
observed gamma-ray and radio emission from the FB. Besides,
accelerated particles are accumulated nearby the $E_{max}$ (see
Eq. (\ref{cut})) forming there an excess of particles which also
leads to a flat spectrum of the nonthermal emission generated by
accelerated electrons. This pile-up effect is similar to that from the analysis of  \citet{pile-up} on electron acceleration by shocks.
These  problems of the model are eliminated by the term of
particle escape with the time $\tau$. The escape term makes the
spectrum steeper as needed for observations.
\item One of the main problems of stochastic acceleration from
a background plasma is (over) heating of the plasma by accelerated
particles as was shown by \citet{wolfe06,east08} because the energy transferred to accelerated particles is quickly dumped into the thermal
plasma. This effect prevents formation of nonthermal spectra. As \citet{chern12} showed, however, the effect of overheating depends on parameters of acceleration, and it is insignificant if
the stochastic acceleration is effective for particles with
high enough momenta $p>p_0$.
 We detemined parameters of acceleration when the  acceleration of electrons in the FB is possible.
\item We described the stochastic (Fermi) acceleration   as a momentum diffusion with the coefficient, $D_F(p)=\alpha p^\varsigma\theta(p-p_0)$, where $p$ is the particle momentum, $p_0$ is a cut-off of the accelertion parameter, and $\alpha$ is the acceleration rate.  The goal of our analysis is to define the model parameters $\alpha$, $\varsigma$, $p_0$ and $\tau$ at which the gamma-ray and radio emission from the FB can be provided by this acceleration mechanism.
\item The value of $p_0$ is determined from the
the  conditions that the  acceleration time $\tau_{acc}\sim1/\alpha$ is
smaller that the time of the plasma heating by the acceleration
particles $\tau_T$. We showed that
for the case of  Bubble plasma the effect of overheating is insignificant if
the stochastic acceleration is effective for particles with a high cut-off
momentum  $p_0\simeq 0.34$ where $p_0$ is given in units of $mc$.
\item The required spectral index of the coefficient of momentum diffusion, $D_F(p)$, is $\varsigma=2$. The effect of particle escape from the acceleration region with the characteristic time $\tau$ is the steepening of the spectrum of acceleration particles. The spectrum required for the observed radio emission from the FB can be obtained if the escape time $\tau=1.1\times 10^{13}$s and the acceleration rate $\alpha \simeq 1.6\times 10^{-14}$s$^{-1}$.  As it is clear from Eq.(\ref{eq_nr}), for $\varsigma=2$ the same effect of steepening can be obtained if partiles lose their energy by adiabatic energy losses inside the FBs instead of escape from there. For the rate of adiabatic losses $dp/dt=-p/3 \nabla\cdot {\bf u}$  the necessary spectral index of accelerated particles can be obtained if 
$1/3 \nabla\cdot {\bf u}=3/\tau$.
 \item  As follows from our numerical
 calculations  the  power supplied by external sources of acceleration in the FB should be about $\sim 4\times 10^{39}$ erg
s$^{-1}$. This
 is lower than $10^{40}$ erg s$^{-1}$ as estimated by \citet{crock11}
 for the energy release provided by the star formation regions in the GC and also is below $10^{41}$ erg s$^{-1}$  as estimated by \citep{cheng} for star
acceleration processes onto the central black hole.
\item In this model the power excess between supplied by external sources and that emitted by electrons in the form of gamma-ray and radio fluxes can be removed from the FBs  either in the form particle escape from the bubbles or by particle interaction with the plasma outflow from the GC region (adiabatic losses).
\item In principle, a physical mechanism for a cut-off $p_0$ in the spectrum of MHD-waves could be wave absorption by cosmic rays.
\item Our investigations showed that for chosen parameters of the background plasma in the FB, the stochastic acceleration is able to provide needed number of high energy electrons in the FB if a set of the acceleration parameters is fixed.  In Table
\ref{tbl-2} we summarized the required parameters of stochastic acceleration needed to reproduce the observed radio and gamma-ray emission from the FB.
\begin{table}[h]
\begin{center}
\caption{Parameters of the model of stochastic acceleration of electrons in the FB.\label{tbl-2}}
\begin{tabular}{ccccccccccccccccc}
&&&&&&&&&&&&&&&&\\
\tableline\tableline
&T (keV)&N (cm$^{-3}$)&H ($\mu$G) & $ \varsigma$  &&& $\alpha  $ (s$^{-1}$) &&& $p_0$ &&& $\tau$ (s) &&&  \\
\tableline
&2&0.01&5 & 2 &&&  $1.6\times 10^{-14}$  &&&0.34    &&& $1.1\times 10^{13}$ &&& \\
\tableline\tableline
\end{tabular}
\end{center}
\end{table}
\end{itemize}

\section*{Acknowledgements}

The authors would like to thank the unknown referee for his careful reading of the text and many critical comments which helped to improve the text of the paper.
KSC is supported by the GRF Grants of the Government of the Hong
Kong SAR under HKU 701013. DOC is supported in parts by  the LPI
Educational-Scientific Complex, RFFI grant 12-02-31648 and 
Dynasty Foundation. DOC and VAD acknowledge support from the RFFI grant 12-02-00005. CMK is supported, in part, by the Taiwan National Science 
Council Grant NSC 102-2112-M-008-019-MY3.
\appendix
\section{Analytical Estimates of the Cut-off Momentum ${\bf p}_0$}\label{appen}

We present here a qualitative analysis as an illustration that MHD wave absorption by CRs may generate
a cut-off in the spectrum of MHD waves.
\citep[for the details of calculations and
references see]{ptus06}. In the stationary case the equation for
spectral energy density of waves, $W(k,t)$  can be written as
\citep[see][]{norman96}
\begin{equation}
\frac{d \Pi(W,k,t)}{dk}= -2\Gamma_{cr}W+\Phi\delta(k-k_0),
\label{Wk}
\end{equation}
where $k$ is the wave-number,  $\Phi$ is energy supplied by the external source at the scale $1/k_0$.
$\Gamma_{cr}$ is the term of wave absorption by CRs, see
\cite{ber90}
\begin{equation}
\Gamma_{cr}(k)=\frac{\pi Z^2
e^2V_A^2}{2kc^2}\int\limits_{p_{res}(k)}^\infty\frac{dp}{p}F(p)\,,
\end{equation}
where $p$ is the particle momentum, $p_{res}(k)=ZeB/ck$, $F(p)$
is the CR distribution. The number density of CRs is $N_{cr}=\int
F(p)dp$ and $F(p)=p^2f(p)$ where $f(p)$ is the distribution
function from Eq. (\ref{eq_nr}).

The  term ${d \Pi(W,k,t)}{dk}$ describes the wave cascade.
Spectrum of MHD turbulence in the interstellar medium is
questionable and usually Kraichnan or Kolmogoroff spectra are
supposed for this medium. Below we assume for simplicity that
there is the Kraichnan spectrum in the FB, then equation for
MHD-waves can be presented in a compact form. For the Kraichnan
spectrum  the wave cascade term is \citep[see][and references
therein]{ptus06}
\begin{equation}
\left(\frac{d \Pi(W,k,t)}{dk}\right)_{Kr}=\frac{d}
{dk}\left[\frac{C\left(k^{3/2}W(k)\right)^{3/2}}{\rho V_A}\right]\,,
\end{equation}
Here  the constant $C\sim 1$, $V_A$ is the Alfven velocity and
$\rho$ is the plasma mass density.

The solution of equation (\ref{Wk}) is given by
\begin{equation}
W(k)=k^{-3/2}\left[k_0^{3/2}W(k_0)-\frac{Z^2e^2B^2V_A}{8Cc^2}\int\limits_{k_0}^kdk_1k_1^{-5/2}
\int\limits_{p_{res}(k_1)}^\infty\frac{F(p)dp}{p}\right]\,,
\label{Wk1}
\end{equation}
where $W(k_0)=\sqrt{\rho V_AS/C}k_0^{-3/2}$.

The coefficient of momentum diffusion $D_p$ is (see
\cite{ber90})
\begin{equation}
D_p(p)=p^2\kappa(p)
\end{equation}
where
\begin{equation}
\kappa(p)=\frac{12\pi V_A^2k_{res}W(k_{res})}{vr_LB^2}.
\label{kappa1}
\end{equation}
Here $B$ is the magnetic field strength and $r_L$ is the particle
Larmor radius $r_L=1/k_{res}$.

From Eqs. (\ref{Wk1}) and (\ref{kappa1}) we have
\begin{equation}
\kappa(p)=\kappa_0(p)\left(1-g\int\limits_p^{p_L}x^{1/2}dx\int\limits_{x}^\infty\,,
dy\frac{F(y)}{y}\right) \label{kappa}
\end{equation}
where
\begin{equation}
\kappa_0(p)=\frac{3\pi}{4}\frac{V_A^2k_L^{3/2}W(k_L)}{vr_LB^2}\,,
\end{equation}
is the diffusion coefficient determined by the Kraichnan spectrum,
the coefficient $g$ is
\begin{equation}
g=\sqrt{\frac{Ze}{\pi c\rho}}\frac{B^{3/2}}{16Ck_0^{3/2}W(k_0)}\,.
\end{equation}

From observations the function $F(p)$ is supposed to be a power-law
with the spectral index between 1.8 to 2.4 (Su et al. 2010). To simplify the calculations
we take $F(p)$ in the form
\begin{equation}
F(p)=\frac{8\times
10^{-12}}{mc} {p}^{-2.25}~~\mbox{cm$^{-3}$mc$^{-1}$}=\frac{\alpha_p}{mc} {p}^{-2.25}
\label{psi}\,.
\end{equation}

The  solution for $\kappa(p)$ can be obtained  in the way similar
to \citet{ptus06}. For the variable $x= {p}^{3/2}$ and the
function $\phi=\frac{\kappa(p)}{\kappa_0(p)}$ we obtain from Eq.
(\ref{kappa})
\begin{equation}
\frac{d^2\phi}{dx^2}=-\frac{4g\sqrt{mc}\alpha_p}{9}\frac{\phi(x)}{x^{2.5}}\,.
\label{phieq}
\end{equation}
Solving Eq. (\ref{phieq}) gives
\begin{equation}
\kappa( {p})=B\kappa_0(p) {p}^{3/2}{J_2(\xi)}
\end{equation}
where $B$ is a constant which can be defined from the boundary
condition that $\kappa( {p})=\kappa_0(p)$ at $p\rightarrow\infty$
\begin{equation}
\xi( {p})=\sqrt{\frac{64}{9}g\alpha_p}(mc)^{1/4} {p}^{-3/8}
\end{equation}
At $\xi=5.14$ the Bessel function $J_2(\xi)=0$. This condition
just determines the cut-off momentum in Eq. (\ref{pcut}).

If we take reasonable parameters for the FB: the
average energy release there $\Phi=10^{39}$ erg s$^{-1}$, the plasma density
$n=10^{-3}$ cm$^{-3}$,
we get $ p_0=\simeq 0.2$.

\begin{figure}[h]
\centering
\includegraphics[width=0.6\textwidth]{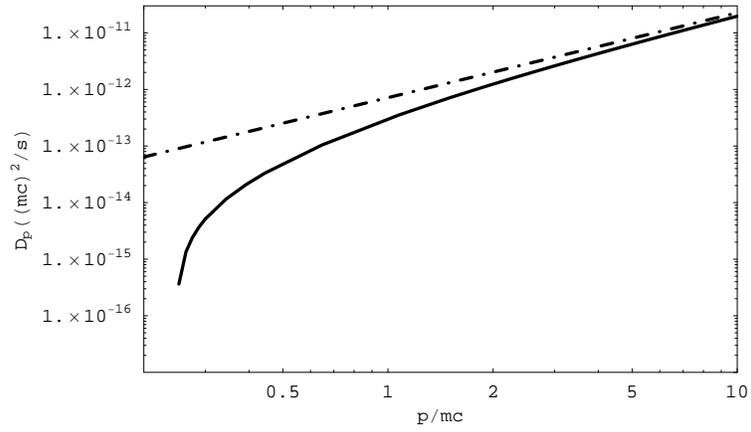}
\caption{The solid line shows the momentum diffusion coefficient
derived for the Bubble parameters when the CR absorption is taken
into account. The dash-dotted line is the results ignoring the CR
absorption.} \label{Dp}
\end{figure}

The momentum diffusion coefficient $D_p$ for the Bubble parameters
is shown in Fig. \ref{Dp} (solid line). For comparison the dash-dotted line
is the diffusion coefficient for the Kraichnan spectrum of
turbulence without CR absorption.

\end{document}